\documentclass[reprint, superscriptaddress,
 amsmath,amssymb,
 aps,
]{revtex4-2}

\usepackage{graphicx}% Include figure files
\usepackage{dcolumn}% Align table columns on decimal point
\usepackage{bm}% bold math
\usepackage{graphicx}                                   % figures
\usepackage{booktabs}                                   % professional tables
\usepackage{cleveref}                                   % references
\crefname{figure}{Fig.}{Figs.}
\Crefname{figure}{Figure}{Figures}
\crefname{equation}{Eq.}{Eqs.}
\Crefname{equation}{Equation}{Equations}
\usepackage{amsmath}                                    % math
\usepackage{siunitx}                                    % units, numbers
\DeclareSIUnit{\bar}{bar}                               % declaring bar as a unit
\newcommand{\CS}{\ensuremath{\mathrm{CS_2}}}            % easy insertion of CS_2 into the text

\begin{document}

\title{Giant Brillouin gain in frozen CS\textsubscript{2} capillaries}

\author{Simon Seiderer}
\thanks{These authors contributed equally to this work.}
\affiliation{Max Planck Institute for the Science of Light, Staudtstr. 2, 91058 Erlangen, Germany}
\affiliation{Department of Physics, Friedrich-Alexander-Universität Erlangen-Nürnberg, Staudtstr. 7, 91058 Erlangen, Germany}

\author{Andreas Geilen}
\thanks{These authors contributed equally to this work.}
\affiliation{Max Planck Institute for the Science of Light, Staudtstr. 2, 91058 Erlangen, Germany}
\affiliation{Department of Physics, Friedrich-Alexander-Universität Erlangen-Nürnberg, Staudtstr. 7, 91058 Erlangen, Germany}

\author{Luan N. Sliwa}
\thanks{These authors contributed equally to this work.}
\affiliation{Max Planck Institute for the Science of Light, Staudtstr. 2, 91058 Erlangen, Germany}
\affiliation{Present address: Department of Physics and Astronomy, Ruprecht-Karls University Heidelberg, Im Neuenheimer Feld 226, 69120 Heidelberg, Germany}

\author{Linqiao Gan}
\affiliation{Max Planck Institute for the Science of Light, Staudtstr. 2, 91058 Erlangen, Germany}
\affiliation{Department of Physics, Friedrich-Alexander-Universität Erlangen-Nürnberg, Staudtstr. 7, 91058 Erlangen, Germany}

\author{Xue Qi}
\affiliation{Leibniz Institute of Photonic Technology, Albert-Einstein-Str. 9, 07745 Jena, Germany}

\author{Mario Chemnitz}
\affiliation{Leibniz Institute of Photonic Technology, Albert-Einstein-Str. 9, 07745 Jena, Germany}
\affiliation{Institute of Applied Optics and Biophysics, Fröbelstieg 1, 07743 Jena, Germany}

\author{Markus A. Schmidt}
\affiliation{Leibniz Institute of Photonic Technology, Albert-Einstein-Str. 9, 07745 Jena, Germany}
\affiliation{Otto Schott Institute of Materials Research, Friedrich Schiller University Jena, Schellingstr. 12, 07743 Jena, Germany}

\author{Birgit Stiller}
\email{birgit.stiller@iop.uni-hannover.de}
\affiliation{Max Planck Institute for the Science of Light, Staudtstr. 2, 91058 Erlangen, Germany}
\affiliation{Department of Physics, Friedrich-Alexander-Universität Erlangen-Nürnberg, Staudtstr. 7, 91058 Erlangen, Germany}
\affiliation{Institute of Photonics, Leibniz University Hannover, Welfengarten 1A, 30167 Hannover, Germany}

\date{\today}

\begin{abstract}
    Stimulated Brillouin-Mandelstam scattering offers exceptional capabilities for photonic signal processing, but current platforms demand performance trade-offs between long interaction lengths, high gain, low optical losses, and practical implementation.
    Here, we demonstrate a novel platform based on the reversible freezing of a carbon disulfide filled liquid-core optical fiber.
    This approach delivers a giant in-fiber Brillouin gain of \SI{434}{\per\watt\per\meter} with a linewidth of \SI{24}{\mega\hertz}, while maintaining low propagation losses in a fully spliced architecture and providing the potential for meter-scale interaction lengths.
    Leveraging this gain, as a proof of principle, we realize an optoacoustic memory operating at sub-nanojoule pulse energies -- more than two orders of magnitude lower than state-of-the-art implementations.
    This power reduction is universal for Brillouin-based fiber applications in general and will enable low-power photonic signal processing and neuromorphic computing, efficient microwave photonics and sensing, as well as in-fiber quantum optomechanics-based technologies.
\end{abstract}

\maketitle

\section{Introduction}
    Since the invention of the laser, nonlinear optics has been a rich and rapidly advancing field.
    A wide range of applications has emerged, including coherent frequency conversion \cite{franken_generation_1961}, optical sensing \cite{maker_study_1965}, supercontinuum generation \cite{alfano_emission_1970}, femtosecond self-mode-locked lasers \cite{spence_60-fsec_1991}, the generation of entangled photon pairs (for example, via spontaneous parametric down-conversion \cite{kwiat_new_1995}, or spontaneous four-wave mixing \cite{takesue_generation_2004}), optical signal processing \cite{doran_nonlinear-optical_1988} and computing \cite{psaltis_optical_1985}.
    In particular, for the latter two applications, one constraint is the high optical power required to access strong nonlinear interactions.
    Identifying high-gain platforms is therefore crucial for a wide range of nonlinear optical effects.
    
    One of the strongest nonlinear effects is stimulated Brillouin-Mandelstam scattering (SBS), which coherently couples two optical waves and one traveling acoustic wave.
    While its magnitude and exponential length-scaling often make Brillouin scattering a limiting factor in telecommunication systems, the underlying optoacoustic coupling can also be advantageously harnessed.
    In the last decades, a variety of Brillouin-based applications have emerged, such as distributed optical sensing \cite{elooz_high-resolution_2014, song_distributed_2006, horiguchi_botda-nondestructive_1989, vallifuoco_high_2025}, optoacoustic signal processing \cite{eggleton_brillouin_2019, stiller_brillouin_2024}, microwave photonics \cite{marpaung_integrated_2019} and on-chip narrow-linewidth lasers \cite{gundavarapu_sub-hertz_2019}.
    Most concepts involving Brillouin scattering benefit directly from a high Brillouin gain because it enhances the energy efficiency and performance of these systems.
    In particular, all-optical signal processing \cite{kalosha_frequency-shifted_2008, preussler_quasi-light-storage_2011} and recently optoacoustic neuromorphic computing approaches \cite{slinkov_all-optical_2025, phang_photonic_2023, becker_optoacoustic_2024} must provide energy-efficient implementations to compete with their very successful electronic counterparts.
    
    Thus, over the recent years various ideas have been pursued to discover optical platforms with high Brillouin gain, such as utilizing new materials, for example chalcogenide glasses \cite{tow_relative_2012, tow_toward_2013, merklein_chip-integrated_2017, kabakova_narrow_2013} and thin-film lithium niobate \cite{rodrigues_cross-polarized_2025, ye_integrated_2025}, or decreasing the effective area of the interaction with suspended waveguides \cite{espinel_brillouin_2017, otterstrom_silicon_2018, gertler_narrowband_2022}, and tapered or micro-structured fibers \cite{dainese_stimulated_2006, tow_relative_2012, tow_toward_2013}.
    This led to enhanced Brillouin applications such as optical memory \cite{merklein_chip-integrated_2017, saffer_brillouin-based_2025}, emitter-receiver schemes \cite{shin_control_2015}, microwave-photonic notch filters \cite{gertler_narrowband_2022} or narrow-linewidth lasers \cite{tow_toward_2013, kabakova_narrow_2013, otterstrom_silicon_2018}.
    Besides these great achievements, the current solutions require a compromise between gain, coupling loss, interaction length, power handling, and manufacturing complexity.

    Recently, liquid-core optical fibers (LiCOF) filled with carbon disulfide (\CS) have emerged as a platform for nonlinear pulse dynamics such as hybrid soliton formation \cite{chemnitz_hybrid_2017, chemnitz_thermodynamic_2018} and supercontinuum generation \cite{chemnitz_hybrid_2017, hofmann_programmable_2025}.
    This system has also been shown to support Brillouin lasing \cite{kieu_brillouin_2013}, and in our recent work we demonstrated its high Brillouin gain factor \cite{geilen_extreme_2023}.
    
    In this work, we freeze LiCOF to realize giant Brillouin gain factors of \SI{434}{\per\meter\per\watt} while preserving low insertion and propagation losses in a low-complexity fiber design with good power handling and straightforward integration into existing fiber architectures.
    To achieve this, we cool the fiber to \SI{77}{\kelvin}, which is well below the melting point of \CS\ (\SI{-112}{\celsius} / \SI{162}{\kelvin}), and analyze the magnitude and tunability of Brillouin scattering in its solid phase. 
    Compared to common higher-Brillouin-gain media, the fabrication and implementation of this platform requires no advanced fiber drawing methods or elaborate equipment.
    The giant Brillouin gain enables the implementation of highly efficient signal processing schemes, which we exemplarily show in an optoacoustic-memory experiment.
    We achieve the storage of optical signals via external optical control pulses with extremely low-energy.
    Leveraging this approach, one can realize Brillouin based optical signal and information processing with orders of magnitude lower power consumption, compared to traditional fiber platforms.
    
    The frozen LiCOF shows reliable performance in terms of high-power handling, practicability, and low-loss connectivity to standard fiber equipment and is straightforward to fabricate \cite{chemnitz_liquid-core_2023} compared to other high-gain Brillouin platforms.
    Moreover, the LiCOF architecture serves as an ideal system for creating and probing exotic material states under precise thermodynamic control \cite{geilen_extreme_2023}, while the high melting point of \CS\ allows using liquid nitrogen to reach the solidification of the fiber core, instead of requiring a cryostat.
    Since the liquid inside the LiCOF is not limited to \CS\ \cite{chemnitz_liquid-core_2023, azhar_nonlinear_2013}, our approach unlocks a completely new type of optical fiber and possible fiber-core materials.

    \begin{figure}
        \centering
        \includegraphics[width=\columnwidth]{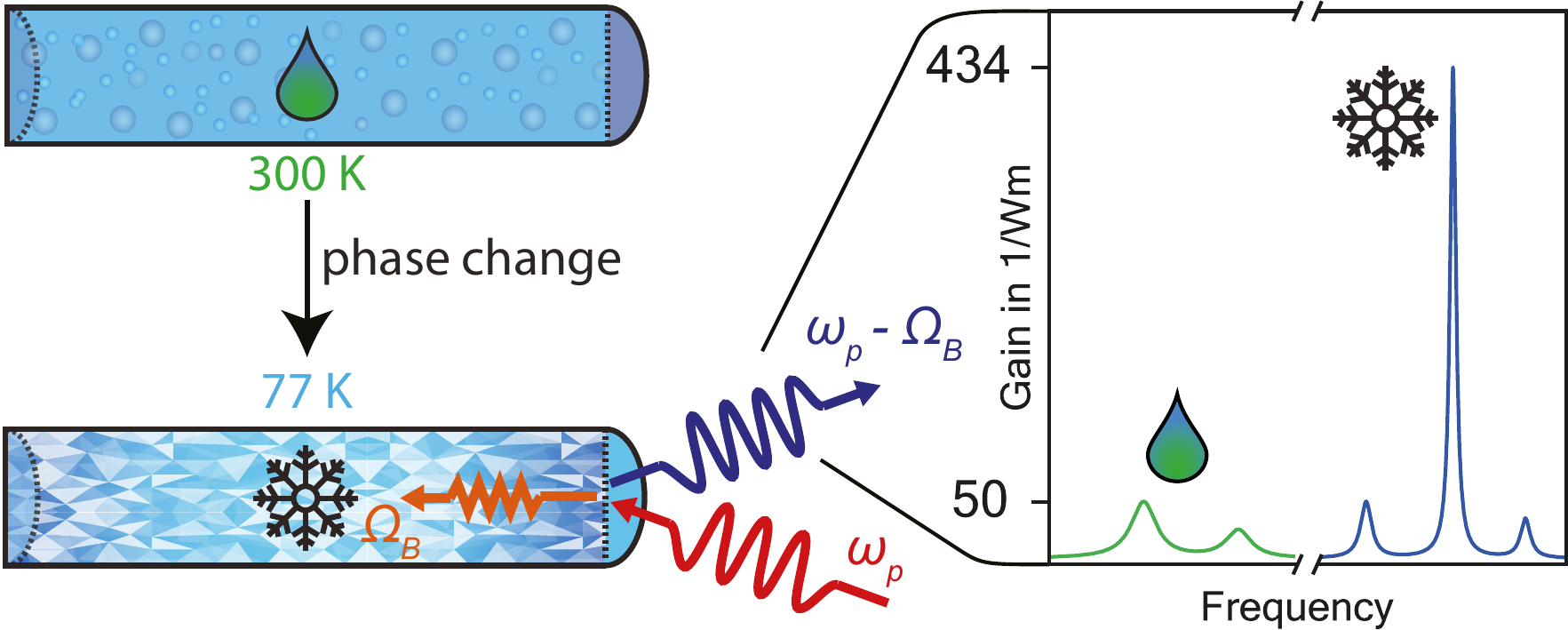}
        \caption{Schematic of Brillouin Scattering in frozen liquid-core optical fiber. A pump photon ($\omega_p$) is back-scattered from an acoustic phonon ($\Omega_B$) resulting in a  back-scattered Stokes downshifted by $\Omega_B$. The back scattered Stokes generated in the frozen section reveals an exceptionally high nonlinearity.}
        \label{fig:Concept}
    \end{figure}
% ############################################################################################
% --------------------------------------------------------------------------------------------
% ############################################################################################
\section{Concept and Fundamentals} \label{sec:Concept}
    \subsection{Thermodynamics of LiCOF}
    We consider a silica capillary filled with liquid \CS\ and fusion spliced to standard single mode fibers, a system which was analyzed in our previous work \cite{geilen_extreme_2023}.
    Such a fully-sealed LiCOF may exhibit several thermodynamic states.
    In the isobaric state at atmospheric pressure, a liquid phase coexists in equilibrium with its vapor.
    In the isochoric state, the core is completely filled with liquid and its volume remains constant.
    In this work, the LiCOF is tailored to operate in the isochoric regime at room temperature.
    
    When such a LiCOF is heated, the thermal expansion of the confined liquid leads to a uniform rise in internal pressure, which is equal along the entire fiber length, while the temperature increase remains local.
    The global pressure increase $\Delta p$ associated with heating a fraction~$L_{\mathrm H}/L_\mathrm{tot}$ of the fiber by a temperature difference~$\Delta T$ is given by 
    \begin{equation}
        \label{eq:DT2Dp}
        \Delta p = \frac{\alpha_{\mathrm V}}{\kappa}\,\frac{L_{\mathrm H}}{L_\mathrm{tot}}\,\Delta T,
    \end{equation}
    where $\alpha_{\mathrm V}=\SI{1.12e-3}{\per\kelvin}$ is the volumetric thermal-expansion coefficient of \CS\ at \SI{20}{\celsius} and $\kappa=\SI{9.38e-10}{\per\pascal}$ is its isothermal compressibility.
    This is valid in the limit $\alpha_{\mathrm V}\,\Delta T \ll 1$.
    Using the same expression, we can understand that when the fiber is cooled, the confined \CS\ contracts and the internal pressure decreases.  
    If the core diameter is sufficiently small, the cohesive forces within the liquid and its adhesion to the capillary wall may prevent the system from relaxing to an isobaric liquid–vapor equilibrium.
    Otherwise, the liquid column breaks down -- a process known as cavitation -- leading to the formation of a cavity filled with \CS\ vapor.

    \begin{figure*}
        \centering
        \includegraphics[width=0.8\textwidth]{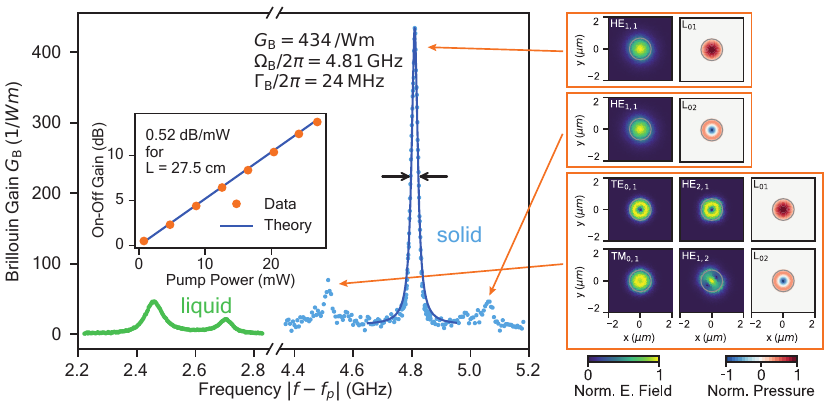}
        \caption{Measurement of Brillouin gain of frozen LiCOF. Brillouin response of liquid (green) and solid (blue) phase demonstrates the up-shift in BFS and increase in Brillouin gain achieved through freezing. Dots correspond to raw measurement data, while the blue line shows the Lorentzian fit of the main peak in the solid phase. The inset shows the exponential increase of the on-off gain with increasing pump power, acquired in a different measurement, and compares it to the theoretical prediction based on the measured $G_B$. On the right, the different features of the frozen LiCOFs spectral response are attributed to different combinations of optical and acoustic modes.
    }
        \label{fig:Gain}
    \end{figure*}

    If cavitation does not occur, the system remains in the metastable state and the liquid sustains large tensile stresses and reaches absolute negative pressures; in Geilen et al. \cite{geilen_extreme_2023} we report values as low as \SI{-300}{\bar}.
    When the temperature of the \CS\ falls below its melting point at \SI{162}{\kelvin}, the liquid freezes. 
    This phase change exhibits interesting characteristics as the higher density of the solid phase is expected to elevate the refractive index of the \CS\ significantly above its ambient value (\SI{1.5885}{} at \SI{1550}{\nano\meter} \cite{kedenburg_linear_2012}).
    
    Maintaining the fiber in the isochoric state is crucial for uninterrupted light transmission -- a strict requirement for any transmission-based application or measurement.
    Although cavitation can be used to switch between isobaric and isochoric states \cite{geilen_extreme_2023}, we suppress it in this work and maintain the isochoric state, allowing us to employ seeded stimulated Brillouin scattering to characterize the frozen LiCOF.
    
    \subsection{Fundamentals of Brillouin-Mandelstam scattering}
    SBS is a coherent, nonlinear coupling between light and acoustic phonons in an optically transparent medium.
    A pump photon interacting with a traveling acoustic wave is scattered into a Stokes photon, which is frequency down-shifted by the Brillouin frequency shift (BFS) and exhibits a Lorentzian-shaped optical spectrum.
    This is schematically shown in \cref{fig:Concept}.
    A much weaker anti-Stokes peak, up-shifted by the BFS, is also present but usually significantly smaller than the Stokes peak under the given conditions \cite{wolff_brillouin_2021}.
    
    The BFS is $\Omega_\mathrm{B}/(2\pi) = 2 n_\mathrm{eff} v_\mathrm{ac} / \lambda$
    and depends on the mode refractive index $n_{\mathrm{eff}}$, the speed of sound $v_{\mathrm{ac}}$ of the acoustic mode and the pump-wavelength $\lambda$.
    When assuming a constant pump power $P_{\mathrm{pump}}$ along a fiber section of length $L$ and neglecting propagation losses (details provided in Supplement~1, Sec.~2), a frequency down-shifted seed with power $P_\mathrm{in}$ is amplified to
    \begin{equation}
        \label{eq:PowerGainEquation}
        P_\mathrm{out} = P_\mathrm{in} \times \mathrm{exp} \left( G_\mathrm{B}  P_{\mathrm{pump}} L \right) \text{,}
    \end{equation}
    with the waveguide-normalized Brillouin gain $G_\mathrm{B}$ \cite{kobyakov_stimulated_2010}.
    The power ratio, $P_\mathrm{out}/P_\mathrm{in}$, yields the amplification of the system and will herein be termed the on-off gain. 
    
    To achieve an efficient Brillouin interaction, one seeks to maximize the Brillouin gain
    $G_\mathrm{B} = g_\mathrm{B}/A_\mathrm{ao}$,
    where $A_\mathrm{ao}$ is the optoacoustic overlap area of the participating optical and acoustic modes.
    Increasing $G_\mathrm{B}$ can either be done by reducing $A_\mathrm{ao}$ or by using a platform with a high intrinsic Brillouin gain coefficient $g_\mathrm{B} = (8 \pi^2 n_\mathrm{eff}^8 p_{12}^2)/(c_0\lambda^3\rho_0\Omega_\mathrm{B}\Gamma_\mathrm{B})$, 
    where $p_{12}$ is the relevant electrostriction tensor element, $c_0$ the speed of light in vacuum, $\rho_0$ the mean material density and $\Gamma_\mathrm{B}$ the Brillouin linewidth \cite{kobyakov_stimulated_2010}.
    As $g_\mathrm{B}$ scales with the eighth power of the effective refractive index $n_\mathrm{eff}$, the increase in refractive index when \CS\ transitions from the liquid to the frozen phase is expected to result in a substantial enhancement of the Brillouin gain.
% ############################################################################################
% --------------------------------------------------------------------------------------------
% ############################################################################################
    \begin{figure*}
        \centering
        \includegraphics[width=0.8\textwidth]{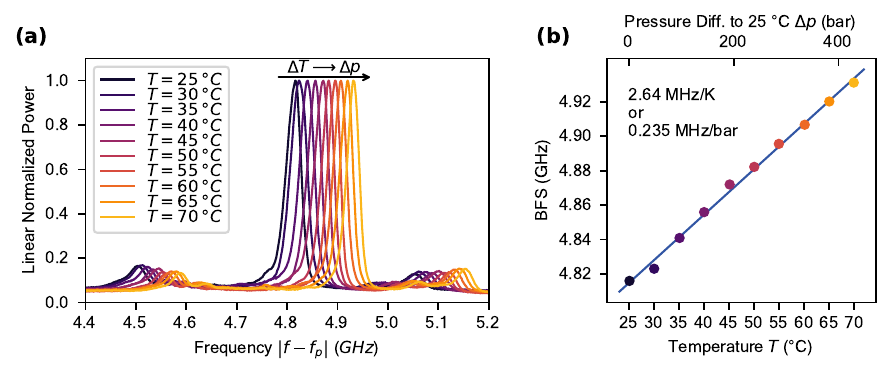}
        \caption{\textbf{(a)} Tuning Brillouin response of frozen LiCOF by heating a fraction $L_\mathrm H$, thus increasing the global pressure, before freezing. \textbf{(b)} BFS of frozen LiCOF as a function of the temperature used for heating $L_\mathrm H$. Pressure difference computed via temperature difference from \SI{25}{\celsius}.}
        \label{fig:TemperatureSweep}
    \end{figure*}

\section{Experimental Methods} \label{sec:ExperimentalMethods}
    The utilized LiCOF is fabricated by filling a silica capillary of total length $L_\mathrm{tot} = \SI{5}{\meter}$ with \CS, a liquid commonly used for nonlinear optics \cite{chemnitz_hybrid_2017, chemnitz_thermodynamic_2018, hofmann_programmable_2025, chemnitz_liquid-core_2023}.
    The fully sealed capillary has a nominal inner diameter of \SI{1.37}{\micro\meter} and is spliced on both sides to single mode optical fibers (SMF) pigtails supported by bridge fibers with ultra high numerical aperture.
    Although the coupling is not optimized for such small core diameters, we measure a relatively low total loss of the sample of \SI{5.6}{\decibel} at ambient conditions. 
    To prevent cavitation while freezing the isochoric fiber, the sample has to retain a positive internal pressure during all experiments.
    
    During our experiments, a segment of length $L_\mathrm{F} = \SI{27.5}{\centi\meter}$ is immersed in liquid nitrogen (\SI{77}{\kelvin}) to freeze the liquid in the fiber core locally \cite{baenziger_crystal_1968}. 
    A different portion of the fiber, $L_\mathrm{H} = \SI{4}{\meter}$, is optionally heated to temperatures of up to \SI{343}{\kelvin} (\SI{70}{\celsius}) to increase the global pressure in the sealed sample via \cref{eq:DT2Dp} (see Supplement~1, Sec.~1 for the schematic setup). 
    This technique is used to determine the influence of the pressure on the freezing conditions.
    A short residual section of the fiber remains at room temperature and the BFS of this section is thus only influenced by the total internal pressure.
    Throughout the experiments, the transmission through the sample is monitored with a laser at \SI{1550}{\nano\meter}.
    
    A pump-probe configuration, adapted from Wolff et al. \cite{wolff_brillouin_2021} and discussed in more detail in Supplement~1, Sec.~1, is employed to characterize the Brillouin response of the LiCOF.
    A continuous-wave (cw) probe at frequency $f$ with an input power $P_\mathrm{in}$ counter-propagates with a cw pump at frequency $f_\mathrm p$ and a power of $P_\mathrm{pump}$.
    The probe is frequency shifted from the pump by $|f - f_\mathrm p|$, which is swept from  \SI{2.2}{\giga\hertz} to \SI{5.2}{\giga\hertz} to span the BFS of the liquid and frozen LiCOF around \SI{2.46}{\giga\hertz} and \SI{4.81}{\giga\hertz}, respectively.
    For every frequency $|f - f_\mathrm p|$, the input power $P_\mathrm{in}$ and amplified output power $P_\mathrm{out}$ of the probe are recorded -- the ratio of these powers $P_\mathrm{out}/P_\mathrm{in}$ defines the on-off gain.
    From the measured on-off gain, the Brillouin gain is obtained using \cref{eq:PowerGainEquation} with the pump power and the interaction length in the liquid and frozen sections, respectively.
    To estimate $P_\mathrm{pump}$ inside the LiCOF from the power sent into the SMF pigtails, we use the total loss of the full sample assembly and assume that it is symmetrically distributed between the input and output sides.
    This tunable pump-probe scheme enables the direct measurement of $G_\mathrm B$ and further offers several advantages: it keeps the interaction in the stimulated regime, provides a high-spectral-resolution access to the Brillouin response, and renders the result independent of probe losses.
% ############################################################################################
% --------------------------------------------------------------------------------------------
% ############################################################################################

    \begin{figure*}
        \centering
        \includegraphics[width=0.8\textwidth]{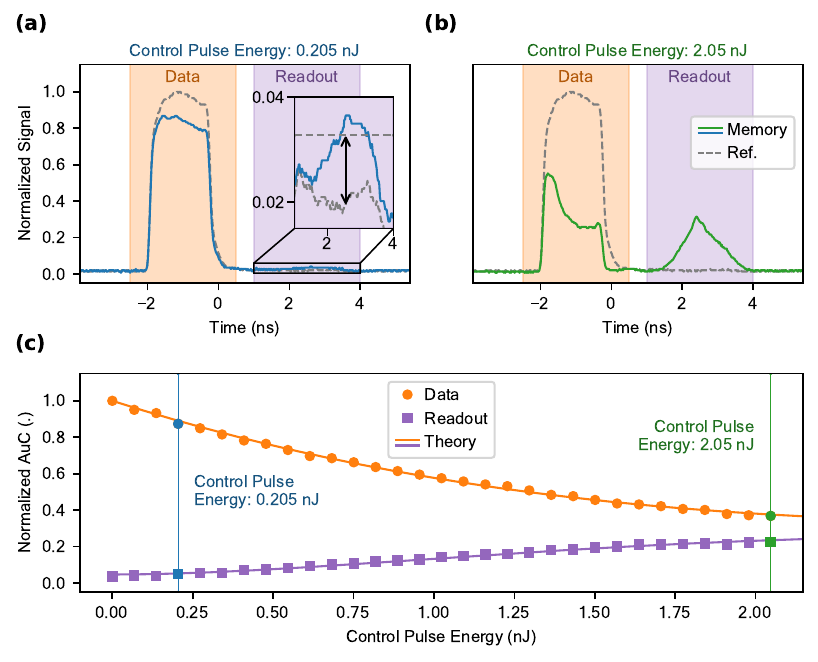}
        \caption{\textbf{(a)} Experimental measurement of the write and read process for a control-pulse energy of \SI{0.205}{\nano\joule}. Gray dashed lines show the reference taken without control pulses. The inset zooms into the readout to show the signal surpassing three standard deviations of the noise above the noise level. \textbf{(b)} Memory process for a control-pulse energy of \SI{2.05}{\nano\joule}. \textbf{(c)} Area under the curve of data and readout pulses computed from the individual measurements for increasing control pulse energies in comparison to the theory.}
        \label{fig:MemoryFig}
    \end{figure*}
\section{Results} \label{sec:Results}

    Upon cooling a segment of the LiCOF, the resulting pressure drop alters both n$_\mathrm{eff}$ and v$_\mathrm{ac}$ thereby modifying the optical and acoustic mode profiles along the entire fiber. 
    This change increases the mode field diameter in the LiCOF, consequently reducing the coupling losses of the splices and yielding a small net increase in optical transmission, persisting after the solidification of the \CS.
    Based on optical time domain reflectometry (OTDR) and Brillouin optical time domain analysis (BOTDA), shown in Supplement~1, Sec.~3, we estimate an effective refractive index $n_\mathrm{eff}$ of the fundamental mode in the frozen LiCOF of \SI{1.94\pm0.21}{} and a low propagation loss $\alpha_\mathrm{dB}$ of \SI{0.20\pm0.08}{\decibel\per\meter}.
    From the increase in refractive index one expects an increase in $g_\mathrm B$ and $\Omega_\mathrm B$, in combination with a tighter confinement of the mode leading to a decrease in $A_\mathrm{ao}$.
    Furthermore, OTDR measurements allow the characterization of the coupling and propagation losses of the system, described in Supplement~1, Sec.~2.
    
    The Brillouin response of the LiCOF reveals the contrast between the liquid phase and frozen sections. 
    In the liquid (green in \cref{fig:Gain}), the resonance between fundamental optical and acoustic modes appears at a Brillouin frequency shift of \SI{2.46}{\giga\hertz} with a linewidth of \SI{71}{\mega\hertz}. 
    In the solid phase (blue), this resonance shifts to $\Omega_\mathrm{B}/2\pi = \SI{4.81}{\giga\hertz}$ with a lowered linewidth of \SI{24}{\mega\hertz}. 
    Additionally, the Brillouin gain of this resonance increases from \SI{47\pm3}{\per\watt\per\meter} to \SI{434\pm22}{\per\watt\per\meter}, representing nearly an order‑of‑magnitude enhancement -- a value only slightly surpassed by a specialty fabricated chalcogenide microstructured fiber taper \cite{tow_toward_2013}.
    However, as stated in \cite{toupin_small_2012}, "it is a challenge to obtain such small-core fiber with low optical losses".
    
    The giant gain measured confirms the predicted tight optical confinement in the frozen LiCOF resulting in a lowered $A_\mathrm{ao}$ and the intrinsically large Brillouin gain coefficient $g_\mathrm B$ of the platform.
    With this outstanding $G_\mathrm B$ we obtain an on-off gain of \SI{6.94}{\decibel} using merely \SI{13.4\pm0.7}{\milli\watt} of pump power over a frozen interaction length of \SI{27.5}{\centi\meter}.
    Additionally, the spectrum of the frozen LiCOF reveals weaker and broader side bands around \SI{4.52}{} and \SI{5.06}{\giga\hertz}.

    With a core refractive index of \SI{2.07}{} we obtain a V-number of \SI{4.1}{} resulting in multi-mode optical guidance of light in the frozen part of the fiber -- also verified by numerical simulations.
    Furthermore, the large acoustic velocity mismatch between \CS\ and silica leads to multiple guided acoustic modes.
    This can explain the smaller peaks in the spectrum.
    While the dominant peak at \SI{4.81}{\giga\hertz} corresponds to a coupling between the fundamental optical and fundamental acoustic mode, the secondary features arise from interactions involving higher-order optical and acoustic modes with increased $A_\mathrm{ao}$.
    The interaction between fundamental optical and acoustic modes yields the lowest $A_\mathrm{ao}$ of \SI{1.28}{\micro\meter\squared}, from which we calculate a Brillouin gain coefficient $g_\mathrm{B}$ of \SI{1.23e-10}{\meter\per\watt}. 
    Further details are provided in Supplement~1, Sec.~4.
    The inset in \cref{fig:Gain} shows the peak on-off gain of the frozen section for various pump powers.
    Based on the measured spectrum, one predicts its on-off gain to scale with \SI{0.52}{\decibel \per \milli \watt}, as verified by the measurement.
    
    Beyond the high Brillouin gain of the frozen LiCOF platform, we demonstrate thermal tunability of its BFS.
    This is achieved by locally heating the liquid section $L_\mathrm{H}$, thereby setting the global pressure in the core before freezing, and subsequently solidifying the section $L_\mathrm{F}$ under these conditions.
    Note that the heated and the solidified sections are spatially disjunct and heating does not influence the temperature in the solidifying section.
    As a result, the BFS measured along $L_\mathrm{F}$ is set by the pressure at solidification and shifts to higher frequencies for higher pressure.
    \Cref{fig:TemperatureSweep}~(a) shows the Brillouin spectra of the frozen segment at different heater temperatures, achieving a total BFS tuning range of \SI{115}{\mega\hertz}.
    \Cref{fig:TemperatureSweep}~(b) shows a linear dependence of the BFS to the heater temperature with a slope of \SI{2.64}{\mega\hertz\per\kelvin}. 
    As with our setup the BFS is only indirectly dependent on the heater temperature via the increase in total pressure before freezing, we convert the local heater temperature to global pressure, finding a slope of \SI{0.28}{\mega\hertz\per\bar}.
    As $\Delta p$ is not only a function of $\Delta T$, but also of $L_\mathrm{H}$, one may increase the tunability with temperature simply by increasing the relative length of the heated fraction $L_\mathrm{H}/L_\mathrm{tot}$.
    In the limit of $L_\mathrm H \rightarrow L_\mathrm{tot}$ we expect a tunability of \SI{3.30}{\mega\hertz\per\kelvin}.
    Additionally, the BFS tuning range could be further increased by increasing the temperature range utilized.
    The pressure-induced increase of the BFS is qualitatively comparable to that reported for liquid \CS\ \cite{geilen_extreme_2023}, though the magnitude is reduced in the solid phase and the shift does not depend on the pressure anymore in the frozen state.
    The thermo-optic contribution to the BFS -- i.e. a down-shift with rising temperature \cite{geilen_extreme_2023} -- could not yet be isolated in the present setup because cooling with liquid nitrogen fixes the frozen segment at a temperature of \SI{77}{\kelvin}. 
    
    Benefiting from its intrinsically high Brillouin gain, the frozen LiCOF provides a suitable platform for energy-efficient applications.
    We show this exemplarily with an optoacoustic memory, where optical information is coherently transferred to acoustic waves at high speeds.
    In the write-process, a "data" and a "control" pulse are sent into the sample.
    These pulses counter-propagate and are separated in frequency by the BFS.
    When they meet in the frozen section of the LiCOF, information is transferred from the data-pulse to the acoustic wave.
    After a storage time $\tau$ the information can be read out with a second control-pulse converting the information from the acoustic wave back into a "readout" pulse.
    For a detailed explanation of Brillouin-based memory principles, refer to \cite{stiller_brillouin_2024}.
    While maximum storage times benefit from the long lifetime of the excited phonons $\tau_\mathrm{ph}$, the high Brillouin gain achieved in the frozen LiCOF platform increases the efficiency of the interaction and thus allows for a low-energy process.
    
    For the implementation of the optoacoustic memory we use data-pulses with a pulse energy of \SI{0.030}{\nano\joule}, pulse durations of $\tau_\mathrm{d/c} = \SI{1.7}{\nano\second}$ for the data and control pulses, respectively, and a storage time of $\tau = \SI{3.0}{\nano\second}$.
    A repetition rate of \SI{5}{\mega \hertz} is used for the experiment, corresponding to a duty cycle of \SI{0.85}{\percent}.
    To assess the performance of our system, the area under the curve (AuC) in a fixed time frame is calculated for the data- and readout-pulses, respectively and normalized to the AuC of the initial data-pulse.
    A key highlight of this implementation is the extremely low control-pulse energy required for effective write- and read-processes.
    As shown in \cref{fig:MemoryFig}~(a), data recovery is achieved with a control-pulse energy of only \SI{0.205}{\nano\joule}, at which the readout signal exceeds the noise floor by three standard deviations.
    Upon increasing the control-pulse energy to \SI{2.05}{\nano\joule}, the best performance is achieved with total write efficiencies of up to \SI{63}{\percent} and read-out efficiencies of up to \SI{23}{\percent}, displayed in \cref{fig:MemoryFig}~(b).
    \Cref{fig:MemoryFig}~(c) shows how the AuC of the depleted data pulses and the recovered readout pulses evolves with the control-pulse energy.
    For comparison, we fit the data with a theoretical model by Zhu et al. \cite{zhu_stored_2007} and reproduce the measured trends with good agreement using four free parameters (see Supplement~1, Sec.~5).

% ############################################################################################
% --------------------------------------------------------------------------------------------
% ############################################################################################
\section{Discussion}
    In this work, we showcased frozen LiCOF, a new type of optical fiber providing a tunable, fully-spliced platform with exceptionally high Brillouin gain, while preserving low coupling and propagation losses.
    The simple design and fabrication procedure of our system is opposed to integrated photonic platforms or chalcogenide fibers, which come with higher insertion and propagation losses and challenges in manufacturing and operation, respectively.
    High gain and low losses are crucial for improving efficiency in existing SBS schemes such as distributed optical sensing, microwave photonics, narrow-linewidth Brillouin lasing or all-optical signal processing and computing.
    
    We have shown that the increase in refractive index -- a direct consequence of the solidification of the \CS\ in the fiber core -- leads to an increase in $G_\mathrm B$, achieving a Brillouin gain more than three orders of magnitude higher compared to that in standard silica fibers.
    % For a comparison with other platforms achieving high $G_\mathrm B$, key values are listed in \cref{tab:gain}.
    Other fiber-based platforms achieving high $G_\mathrm{B}$ include carbon dioxide filled hollow core fibers with $\SI{1.68}{\per\watt\per\meter}$ \cite{yang_intense_2020}, highly nonlinear fibers with $\SI{3.0}{\per\watt\per\meter}$ \cite{deroh_comparative_2020}, silica nanofibers with $\SI{15}{\per\watt\per\meter}$ \cite{zerbib_stimulated_2024} and microstructured and tapered chalcogenide fibers with $\SI{550}{\per\watt\per\meter}$ \cite{tow_toward_2013}.
    Notably, only tapered chalcogenide photonic crystal fibers can reach gain values in the same order of magnitude, although this comes at the cost of free-space coupling, higher propagation losses (\SI{0.65}{\decibel\per\meter} \cite{tow_toward_2013}) and very challenging fabrication \cite{toupin_small_2012}, whereas our fully-spliced architecture simplifies the integration with existing telecommunication devices.
    
    Further optimization of the total amplification is possible by either decreasing the core size or by increasing the length of the frozen section.
    Based on numerical simulations (see Supplement~1, Sec.~6) we expect a possible reduction in $A_\mathrm{ao}$ of up to $\times \SI{1.5}{}$ for a core diameter of \SI{0.8}{\micro\meter}, leading to a possible Brillouin gain of approximately \SI{650}{\per\watt\per\meter}.
    As current samples support the guidance of multiple optical modes in the frozen section, a lower core size would not only lead to a lowered $A_\mathrm{ao}$ but can also lead to single-mode propagation of light in the entire LiCOF.
    The maximum amplification for a long fiber is determined by the propagation losses, captured by using the effective length $L_\mathrm{eff} = \left(1 - e^{-\alpha L} \right)/\alpha$.
    For $L \gg 1/\alpha$ this expression converges to $1/\alpha$.
    With an $\alpha_\mathrm{dB}$ of \SI{0.20\pm0.08}{\decibel\per\meter} we calculate a maximum possible slope efficiency of \SI{42}{\decibel\per\milli\watt} for $L_\mathrm{eff} = 1/\alpha$, compared to the \SI{0.52}{\decibel\per\milli\watt} obtained with an interaction length of \SI{27.5}{\centi\meter} in this work.
    A comparison of the slope efficiency as a function of the interaction length for different platforms is given in Supplement~1, Sec.~6.
    
    We showed that the Brillouin response of the frozen LiCOF can be tuned with pressure, by heating part of the liquid before freezing. While not directly enabling temperature sensing, this approach is especially useful in tuning microwave signal-generation \cite{wang_photonic_2012} or Brillouin lasers \cite{popov_electrically_2017} and enables fundamental studies of the freezing properties at high internal pressure values and low temperatures \cite{geilen_high-speed_2024}
    
    Leveraging the Brillouin interaction, several approaches to all-optical computing like optoacoustic memory \cite{stiller_brillouin_2024}, a nonlinear activation function \cite{slinkov_all-optical_2025}, or an optoacoustic recurrent operator \cite{becker_optoacoustic_2024} have already been demonstrated.
    The frozen LiCOF provides a high gain platform that increases the efficiency of these schemes and paves the way towards coherent optoacoustic building blocks for neuromorphic computing with low power consumption.
    We prove the feasibility of our platform by implementing an optical memory using more than two orders of magnitude less energy, than needed with conventional fibers.
    This is qualified by analyzing the pulse area, a criterion for the efficiency of the optical storage, $\Theta = \tau_\mathrm c\sqrt{c G_\mathrm B P_\mathrm c / 8 n_\mathrm{eff} \tau_\mathrm{ph}}$, taken from Zhu et al. \cite{zhu_stored_2007} for rectangular pulses.
    The high Brillouin gain significantly enhances $\Theta$, highlighting the efficiency of the frozen LiCOF platform for energy-efficient Brillouin memory applications.
    Given the parameters of our system, we calculate $\Theta$ for the exemplary traces shown in \cref{fig:MemoryFig}~(a) and (b) to be \SI{0.42}{} and \SI{1.32}{} (Supplement~1, Sec.~5), respectively, where the latter is close to the point of optimal memory efficiency at $\Theta=\pi/2$.
    Considering the conditions for optimal depletion \cite{zhu_stored_2007,stiller_cross_2019}, the results indicate that further optimizing the experimental conditions such as spatial overlap, pulse shape optimization, and control pulse shortening will further increase the memory efficiency in the present system.
    In comparison to the parameters of the highly nonlinear fiber in a cryostat \cite{saffer_brillouin-based_2025}, the frozen LiCOF requires \SI{150}{} times less pump power in order to reach the same $\Theta$. 
    As this reduction in power is generalizable to other similar concepts such as optoacoustic recurrent operators \cite{becker_optoacoustic_2024} or all-optical non-linear activation functions \cite{slinkov_all-optical_2025}, the frozen LiCOF enables the realization of optoacoustic building blocks for optical neural networks with orders of magnitude lower power consumption, compared to traditional platforms \cite{saffer_brillouin-based_2025}.
    Furthermore, this platform might open opportunities for other Brillouin-based applications such as energy-efficient microwave filters \cite{marpaung_integrated_2019} and delay lines, high-resolution Brillouin fiber sensing \cite{elooz_high-resolution_2014, song_distributed_2006, horiguchi_botda-nondestructive_1989, vallifuoco_high_2025}, low-threshold Brillouin fiber lasers \cite{tow_toward_2013, kabakova_narrow_2013, otterstrom_silicon_2018} and Brillouin-based non-Hermitian photonics with exceptional points \cite{bergman_observation_2021}.

    The high Brillouin gain of frozen LiCOF may also pave the way towards quantum applications of Brillouin scattering.
    In particular, it could provide a fully spliced platform for Brillouin cooling towards the acoustic ground state with performances comparable to chalcogenide fibers \cite{fischer_brillouinmandelstam_2026}.
    Operation in the strong-coupling regime is relevant for quantum signal processing due to its coherent transfer of information.
    Optoacoustic strong coupling between optical photons and traveling acoustic phonons has recently been reported in a cavity-less system, specifically in a cryogenic HNLF at \SI{4}{\kelvin} \cite{blazquez_martinez_cavity-less_2025}. 
    As a figure of merit we calculate the cooperativity \cite{aspelmeyer_cavity_2014} of our system using the definitions of coupling strength $g_\mathrm{om}$ and optical dissipation rate $\gamma_\mathrm{opt}$ in Ref. \cite{blazquez_martinez_cavity-less_2025}, and obtain a cooperativity of \SI{102}{} for the frozen LiCOF at $P_\mathrm{pump} = \SI{50}{\milli\watt}$ -- an indication of the potential to reach strong coupling in our system.
    
    Finally, the frozen LiCOF concept can be generalized and applied to other liquids instead of \CS\ \cite{chemnitz_liquid-core_2023}, allowing the characterization of the liquid and solid phase of a wide range of materials at hard-to-reach conditions.
    Brillouin scattering allows for the simultaneous access to its optical and acoustic properties in the liquid and solid state and the fully spliced architecture allows for the straight-forward use of other methods commonly used for characterizing optical fibers.
    
    With SBS being a widely used nonlinear effect in distributed optical sensing, microwave photonics, narrow-linewidth lasers and all-optical signal processing and computing, our work provides a simple to be realized device with exceptional Brillouin gain and straight forward adoption into existing schemes, directly benefiting most Brillouin-based applications by greatly increasing performance and lowering power requirements.
% ############################################################################################
% --------------------------------------------------------------------------------------------
% ############################################################################################

\vspace{0.8\baselineskip}
\section*{Funding}
    S.S., A.G., L.S., L.G. and B.S. acknowledge funding from the Eruopean Unions's ERC Consolidator Grant "Sound-PC" (101170362), the Deutsche Forschungsgemeinschaft (DFG) under grants STI-792/71 and STI-792/1-1, and the Max Planck Society through the Independent Max Planck Research Groups scheme. M.A.S. acknowledges funding from the DFG under grants SCHM2655/3-2 and SCHM2655/23-1. M.C. acknowledges funding by the Carl-Zeiss-Stiftung (CZS) through the Nexus program (P2021-05-025).
\section*{Acknowledgments}
    The authors would like to thank V. Marathan Kovil and R. Keding for helpful and supporting discussion, M. Frosz for lending of the OTDR and the mechanical workshop staff of the MPL, specifically M. Schwab for manufacturing individual components.
\section*{Disclosures}
    The authors M.C. and M.A.S. hold patents on the fabrication and the locally distributed temperature control of liquid-core optical fibers. The authors declare no conflict of interest.
\section*{Data Availability Statement}
    Data underlying the results presented in this paper are not publicly available at this time but may be obtained from the authors upon reasonable request.
\section*{Supplemental document}
    See Supplement 1 for supporting content.

\bibliography{ref}

\clearpage
\onecolumngrid

\section*{Supplementary Information}
\setcounter{section}{0}
\setcounter{subsection}{0}
\setcounter{figure}{0}
\setcounter{table}{0}
\setcounter{equation}{0}
\renewcommand{\thefigure}{S\arabic{figure}}
\renewcommand{\thetable}{S\arabic{table}}
\renewcommand{\theequation}{S\arabic{equation}}

\section{Experimental Setup}
\begin{figure}[b]
    \centering
    \includegraphics[width=0.8\textwidth]{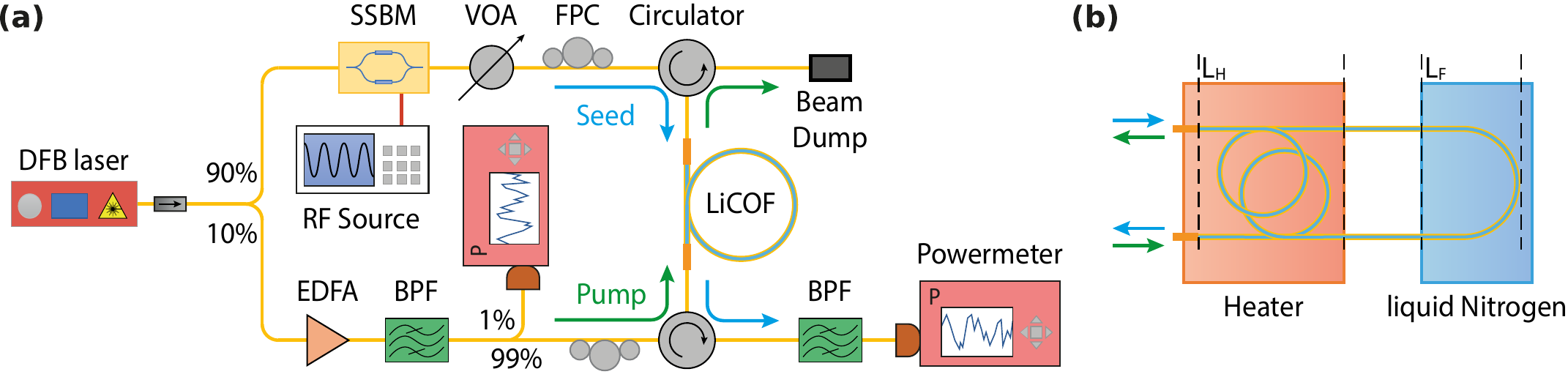}
    \caption{Schematic of Setups. \textbf{(a)} Setup used for gain measurements. DFB laser: distributed feedback laser, SSBM: single-sideband modulator, RF Source: radio-frequency source, VOA: variable optical attenuator, FPC: fiber polarization controller, Circulator: fiber-integrated optical circulator, EDFA: erbium-doped fiber amplifier, BPF: band-pass filter. \textbf{(b)} Experiment configuration with connections to optical setup. Heater to increase pressure, liquid nitrogen to freeze LiCOF. }
    \label{fig:setup}
\end{figure}

The experimental setup used for the gain measurements (see Fig.~2 of the main text) is shown schematically in \cref{fig:setup}~(a). The output of a distributed feedback laser (DFB) operating at a wavelength of \SI{1.55}{\micro\meter} is split into a pump and seed arm.
In the seed arm a single-sideband-modulator (SSBM) shifts the frequency of the laser, a variable optical attenuator (VOA) allows for the control of the seed power and a circulator guides the seed into the far end of the sample.
In the pump arm the laser is amplified by an Erbium-doped fiber amplifier (EDFA), its amplified spontaneous emission is filtered with a band-pass filter (BPF), and launched into the fiber, in opposite direction to the seed.
Fiber polarization controllers (FPC) in each arm help aligning the polarizations of pump- and seed light for an efficient interaction in the sample.
A powermeter continuously monitors the pump power coupled into the LiCOF.
After the back-reflection and spontaneous Brillouin scattering from the liquid sections are removed with a narrow band-pass filter, the amplified seed power is measured using a power meter.
Sweeping the seed frequency via the SSBM thus enables the direct measurement of a spectrum of $G_\mathrm{B}$.
By utilizing one common laser to create the pump and probe pulses respectively and introducing the frequency-shift through an SSBM, slow drifting of the laser frequency do not affect the measurement.

For the temperature sweep shown in Fig.~3 of the main manuscript, a faster measurement technique was used.
Instead of using the split-off arm as a seed, it uses a local oscillator (LO) for heterodyne detection.
The far end of the fiber remains open and the pump injected into the sample leads to noise-initiated Brillouin scattering.
The back-reflected signal can be combined with the LO on a photodiode, and with aligned polarizations their beating can be measured on an electrical spectrum analyzer.
Due to the nature of this approach, measurements are quicker, while the amplitude of the measured signal does not directly yield $G_\mathrm{B}$ anymore.
In order to heat the section $L_\mathrm{H}$ of the fiber before freezing a spatially disjunct section $L_\mathrm{F}$, as described in the main manuscript, we use a setup schematically shown in \cref{fig:setup}~(b).

For the memory-measurements, both the pump and the seed arm, commonly called data and control, respectively, are equipped with intensity modulators and EDFAs, to enable pulsed operation.
The detection is done in the time domain using a photodiode and oscilloscope.
For a more detailed description of the memory process, refer to \cite{stiller_brillouin_2024}.

\section{Optical Time-Domain Reflectometry}
The refractive index is a crucial parameter for modeling optical fibers.
Although the structure and density of frozen carbon disulfide have been characterized \cite{baenziger_crystal_1968}, its refractive index has not been reported.
Additionally, the applicability of these bulk measurements is limited due to the incomplete knowledge about the freezing process within the frozen LiCOF.
Thus, measuring the refractive index is of great interest.

\begin{figure}
    \centering
    \includegraphics[width=0.8\textwidth]{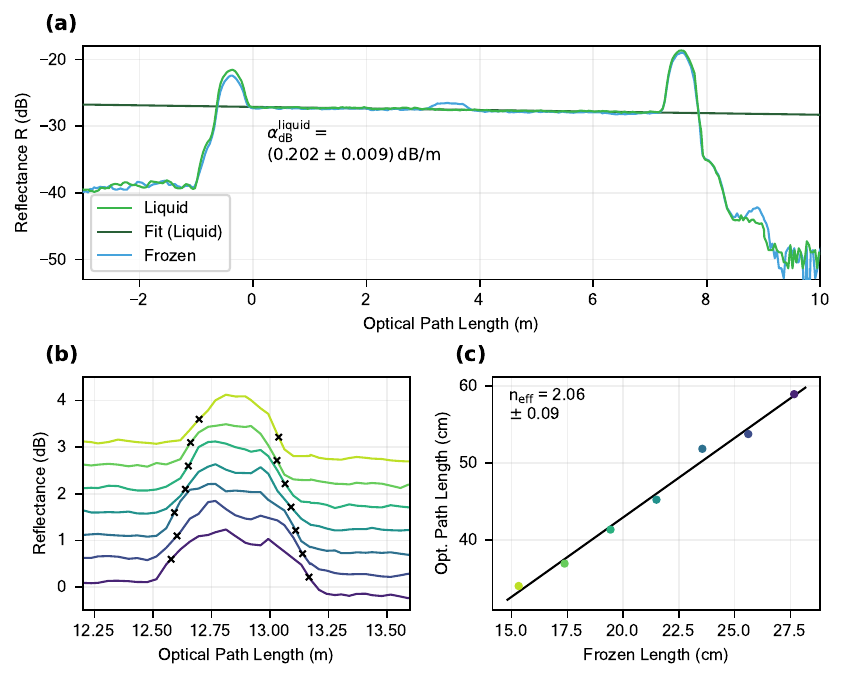}
    \caption{OTDR measurements of the frozen LiCOF. \textbf{(a)} LiCOF before (green) and after (blue) freezing. Two peaks originate from the coupling sections, the liquid baseline is fitted (dark green) to reveal the propagation loss of the fundamental mode inside the liquid region. The frozen trace exhibits a back-reflection at the frozen section, while showing no changes in the liquid section. \textbf{(b)} Back-reflection for different lengths of frozen fiber. \textbf{(c)} Linear fit of the optical path length of the back-reflections from (c) as a function of frozen length reveals an estimate for the effective refractive index.}
    \label{fig:RefractiveIndexFromOtdr}
\end{figure}

A basic measurement technique that allows the measurement of the optical path length of a sample, and thus the calculation of its refractive index via the physical length, is optical time domain reflectometry (OTDR).
Here a pulse is launched into the fiber and the back-scattered signal is recorded as a function of time.
\Cref{fig:RefractiveIndexFromOtdr}~(a) shows the raw traces of the LiCOF in the liquid state and with a frozen section, respectively.
Here the data is already converted to show a reflected signal as a function of the optical path length (OPL).
The spatial resolution depends on the pulse width, typically around $\SI{2}{\nano \second}$ for commercial systems.
Via the OPL between two events (for example the splices and bridge fibers connecting the LiCOF to standard single-mode fibers) one can measure the OPL of a fiber region and subsequently the effective refractive index of the optical mode using the physical length.

Additionally, OTDR can be used to characterize the losses of a system.
\Cref{fig:RefractiveIndexFromOtdr}~(a) shows the reflectance of the LiCOF in the liquid state (green) and with a section frozen (blue).
The slope of a linear fit (dark green) on the liquid trace between the splices, visible as peaks in the reflectance, yields the propagation loss through the liquid sections of the LiCOF.
We show an exemplary measurement and calculate the propagation loss to be \SI{0.202\pm0.009}{\decibel\per\meter} based on five consecutive measurements.

Averaging four independent measurements, we estimate the propagation loss of the frozen section to be \SI{0.20\pm0.08}{\decibel\per\meter}, where the larger standard deviation results from the limited spatial resolution of the OTDR.
In the main manuscript we correct the pump power by half of the total transmission loss through the fiber, effectively implementing the discussed losses into our calculation without the need to distinguish between insertion and propagation losses.

In order to calculate the effective refractive index of the frozen section we need its OPL.
As it is difficult to accurately define this in \cref{fig:RefractiveIndexFromOtdr}~(a), we sweep the frozen length and define the OPL as the distance between the points, where the signal crosses a certain threshold above the noise floor as shown in \cref{fig:RefractiveIndexFromOtdr}~(b).
Now, when fitting the OPL versus frozen length in \cref{fig:RefractiveIndexFromOtdr}~(c), the specific definition of the threshold value arbitrarily changes the offset of the fit.
The slope of this fit reveals an effective refractive index of \SI{2.06\pm0.09}{}.
Note that this value is not the final effective refractive index, as an additional measurement technique is presented in the following section.

\section{Brillouin Optical Time-Domain Analysis}
\begin{figure}
    \centering
    \includegraphics[]{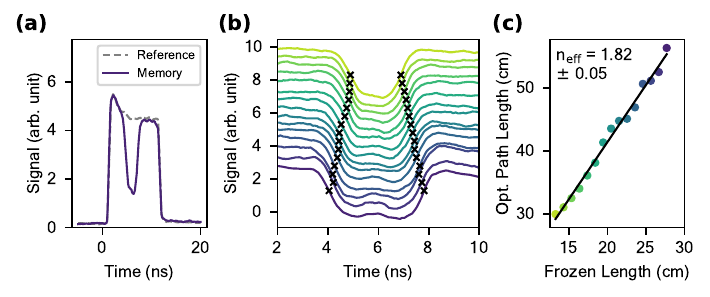}
    \caption{Measurement of the refractive index using a version of BOTDA. \textbf{(a)} Depletion of a control pulse by interaction with a data pulse. \textbf{(b)} Depletion of the control broadens with increasing frozen length. Width of the depletion in time-domain allows for the calculation of optical path length of the frozen section. \textbf{(c)} Fitting the optical path length of the frozen section onto the physical length yields the refractive index of the solid \CS.}
    \label{fig:RefractiveIndexFromMemory}
\end{figure}

In order, to have a second measurement of the refractive index the measurement setup of the optoacoustic memory can be repurposed to do Brillouin optical time domain analysis (BOTDA), where a cw probe is amplified by a short pump pulse and evaluated in the time domain \cite{lu_distributed_2019}.
Equivalently, we use a long pump pulse at frequency $f_\mathrm p$ and a short probe pulse shifted to match the BFS of the frozen section of the LiCOF and measure the depletion of the pump pulse as shown in \cref{fig:RefractiveIndexFromMemory}~(a).
As the pump pulse is longer in space compared to the frozen section, while the length of the probe pulse is shorter, the width of the depletion is proportional to the length of the frozen section.
This allows to convert the depletion-width in the time-domain, shown in \cref{fig:RefractiveIndexFromMemory}~(b), to the optical path length as a linear function of the frozen length, the slope of which (\cref{fig:RefractiveIndexFromMemory}~(c)) yields the effective refractive index of the frozen LiCOF.
This method yields $n_\mathrm{eff}$ = \SI{1.82 \pm 0.05}{}, where the error is derived from the standard error of the fit due to fluctuations in the width of the depletion.
The errors in the determination of the physical frozen length are considered small and are therefore neglected.
Due to the discrepancy with the previously shown OTDR measurements, we report a conservative effective index of $n_\mathrm{eff}$ = \SI{1.94\pm0.21}{} for the fundamental optical mode by averaging the two results and assuming that the majority of the optical power propagates in the fundamental optical mode of the frozen LiCOF.
We attribute the uncertainty to the limited spatial resolution of the two measurements, each utilizing pulse widths in the same order of magnitude as the frozen LiCOF physical length.
Nevertheless, this value confirms that the frozen LiCOF supports multi-mode light propagation (V $>$ \SI{2.405}{} for core refractive indices $n>$ \SI{1.72}{}).
The precision could be improved in future work by increasing the frozen length or employing Optical frequency-domain reflectometry (OFDR), where commercially available systems reach sub-mm resolution \cite{lu_distributed_2019}.

Note that measurements can only yield the effective refractive index of the fiber under test.
In the next section, numerical simulations allow us to extract the refractive index of frozen \CS\ using the known refractive index of the cladding material.

\section{Numerical Simulations}
\begin{figure}
    \centering
    \includegraphics[width=0.8\textwidth]{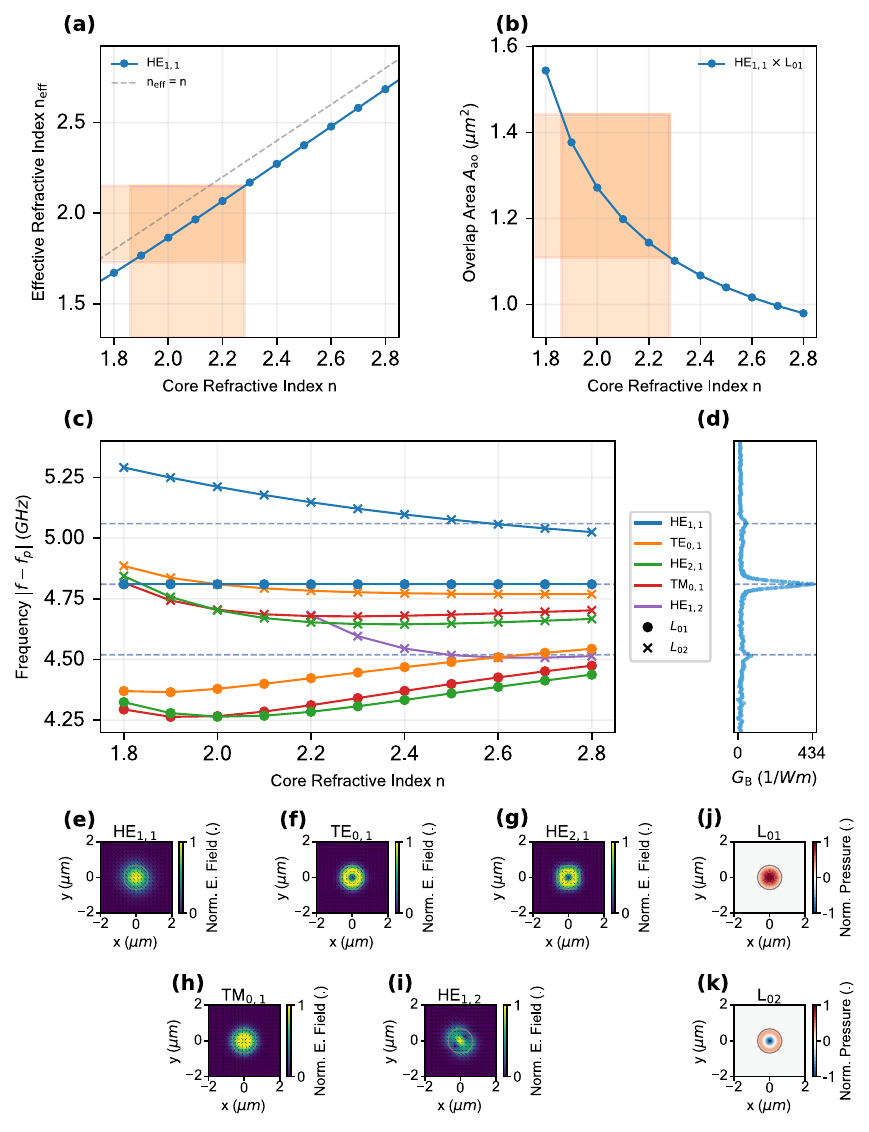}
    \caption{Simulation of the frozen LiCOF platform for a core diameter of \SI{1.37}{\micro\meter}. \textbf{(a)} Relation between refractive index of the core and effective refractive index of the fundamental optical mode of a step-index fiber with silica cladding. Orange rectangles mark the measured values for $n_\mathrm{eff}$ and the corresponding theoretical values for n. \textbf{(b)} Acousto-optic overlap area as a function of n for the interaction between fundamental optical and fundamental acoustic mode. Orange rectangles mark the theoretical values for n obtained from a and the corresponding values for $A_\mathrm{ao}$. \textbf{(c)} Simulated Brillouin frequency shift of different combinations of modes for different n. Horizontal lines mark the maxima of the measured peaks of the main manuscript. \textbf{(d)} Spectrum shown in the main manuscript for comparison to (c). \textbf{(e-i)} Examples for optical modes of the frozen LiCOF. \textbf{(j, k)} Examples for acoustic modes of the frozen LiCOF.
    \label{fig:SimulationFigure}}
\end{figure}

Due to the discrepancies between different measurements we use a range of refractive indices for the numerical simulations.
First, only the optical modes for a step-index fiber with core diameter $\SI{1.37}{\micro \meter}$ are simulated.
The cladding refractive index is kept at \SI{1.444}{}, while the core refractive index n is swept from \SI{1.8}{} to \SI{2.8}{}.
The corresponding mode profiles for a selection of optical modes are shown in \cref{fig:SimulationFigure}~(e - h).
\Cref{fig:SimulationFigure}~(a) shows the effective refractive index of the fundamental mode (e) as a function of the core refractive index.
From our measurements we obtained $n_\mathrm{eff}$ = \SI{1.94\pm0.21}{}, marked on the y-axis of \cref{fig:SimulationFigure}~(a).
With the simulation we can convert from the effective refractive index of a mode to the actual refractive index of the core material, yielding \SI{2.07\pm0.21}{} for the refractive index of frozen \CS.

Next, the phase-matching condition of Brillouin scattering is leveraged to search for acoustic modes with matching wavevector.
One can solve for multiple longitudinal acoustic modes, whereas only the first two orders shown in \cref{fig:SimulationFigure}~(j, k) are of interest here.
The simulations not only yield the Brillouin frequency shift (the eigenfrequency of the respective acoustic mode), but also allows for the calculation of the optoacoustic overlap following \cite{kobyakov_stimulated_2010}.
In general we filter optical and acoustic modes for those that participate in combinations exhibiting an $A_\mathrm{ao}<\SI{1e4}{\micro\meter^2}$.
As we do not have any information on the acoustic velocity $v_\mathrm{ac}$ in frozen \CS\ we calculate this parameter to fit our measurements.
In explicit, for each n we change $v_\mathrm{ac}$ to achieve a BFS of \SI{4.81}{\giga\hertz} for the interaction between the fundamental optical and acoustic mode $HE_{1,1} \times L_{01}$.

\Cref{fig:SimulationFigure}~(b) shows $A_\mathrm{ao}$ for the interaction of fundamental optical and acoustic mode.
This yields \SI{1.28\pm0.17}{\micro \meter^2} from our refractive index measurements, which can be utilized in combination with $G_\mathrm B$ to estimate the Brillouin gain coefficient $g_\mathrm B$ to be \SI{1.23 \pm 0.05 e-10}{\meter\per\watt}.
Higher order combinations are not shown here due to their orders of magnitudes higher $A_\mathrm{ao}$.

Examining the BFS for different optical and acoustic mode combinations allows to look deeper into the physics that eventually yield the final spectral response (\cref{fig:SimulationFigure}~(d)).
While the $HE_{1,1} \times L_{01}$, as the combination with highest $A_\mathrm{ao}$, remains fixed at \SI{4.81}{\giga\hertz}, we can observe $HE_{1,1} \times L_{02}$ matching the smaller spectral feature around \SI{5.06}{\giga\hertz} for $n \approx$ \SI{2.6}{}.
We attribute this discrepancy between measurement and simulation in the refractive index of the frozen \CS\ to deviations in the measured core diameter and assuming an amorphous core structure in the simulation.
Note that we do not show the $A_\mathrm{ao}$ for any higher order combination here, due to the difficulty in comparing these to the measurement without the knowledge of which optical mode carries which fraction of the pump power sent to the sample.
Thus, comparing different $A_\mathrm{ao}$ only works relatively between different acoustic modes combined with the same optical mode.
For comparison to $HE_{1,1} \times L_{01}$, the simulated $A_\mathrm{ao}$ of $HE_{1,1} \times L_{02}$ is about \SI{20}{} times lower, roughly matching our measurements.

Explaining the features around \SI{4.52}{\giga\hertz} is more complex.
At first glance the combinations $(TE_{0,1}; HE_{2,1}; TM_{0,1}) \times L_{01}$ seem to be promising candidates.
However, with the pump power arriving at the frozen section in the $HE_{1,1}$ mode of the LiCOF, which is single-mode at room temperature, ideally no light should be coupled into these optical modes.
This is not the case for $HE_{1,2}\times L_{02}$, also fitting the experimental data nicely, although power in the $HE_{1,2}$-mode should also make the combination $HE_{1,2}\times L_{01}$ with higher $A_\mathrm{ao}$ detectable around \SIrange{3.5}{3.9}{\giga\hertz}.
We propose to examine this frequency region more closely in future measurements in order to better understand the behavior of the frozen LiCOF and possibly confirm the origin of these features.
This knowledge can later be used to optimize parameters like the core diameter to tailor the spectral response of the frozen LiCOF.

\section{Optoacoustic Memory Theory}

Following \cite{zhu_stored_2007} the write and read efficiency of the optoacoustic memory depend on the area of the control pulses defined as
\begin{equation}
    \Theta = \tau_\mathrm{c}\sqrt{G_\mathrm{B} P_\mathrm{c} c / (8 n_\mathrm{eff} \tau_\mathrm{ph})}
\end{equation}
for rectangular pulses.
Control pulses with various peak powers $P_\mathrm{c}$ and pulse widths $\tau_\mathrm c = \SI{1.7}{\nano\second}$ are utilized.
Further using $G_\mathrm{B} = \SI{434}{\per\watt\per\meter}$, $\tau_\mathrm{ph} = 1/(2 \pi \times \SI{24}{\mega\hertz}) = \SI{6.6}{\nano\second}$ and $n_\mathrm{eff} = \SI{1.94}{}$ we obtain $\Theta = 1.92 \sqrt{P_\mathrm{c}/\SI{}{\watt}}$.
In comparison to \cite{saffer_brillouin-based_2025} reaching $\Theta = 0.157 \sqrt{P_\mathrm{c}/\SI{}{\watt}}$ the frozen LiCOF needs $\times 150$ less power in order to achieve the same $\Theta$.

\begin{figure}
    \centering
    \includegraphics[width=0.9\linewidth]{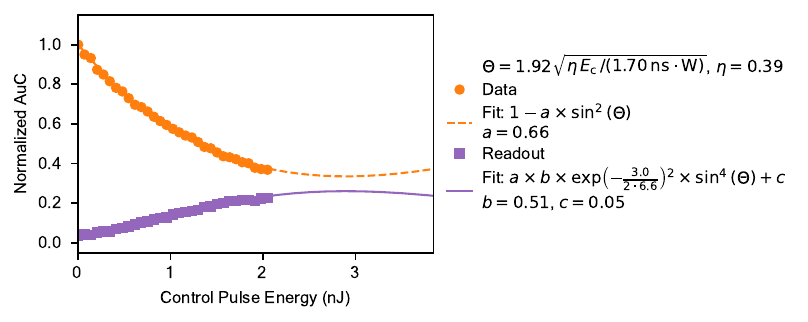}
    \caption{Measured AuC for depleted data and retrieved read-out pulses fitted with theoretical model.
    \label{fig:MemoryFit}}
\end{figure}

The power of the induced acoustic excitation is $a\times\mathrm{sin}(\Theta)^2$ \cite{zhu_stored_2007}, thus the depleted data pulse energy (orange in Supplement~1, Fig.~4~(c)) follows 
\begin{equation}
    E_\mathrm{data} = 1 - a \times \mathrm{cos}(\Theta)^2 \text{,}
\end{equation}
with the fitting parameter $a$ which accounts for an imperfect interaction between data and write pulses.
After an acoustic decay, dependent on the storage time $\tau = \SI{3.0}{\nano\second}$, the energy of the read-out follows \cite{zhu_stored_2007}
\begin{equation}
    E_\mathrm{read\text{-}out} = (a\times\mathrm{sin}(\Theta)^2) \times \mathrm{exp}(-\tau/(2 \tau_\mathrm{ph}))^2 \times (b\times\mathrm{sin}(\Theta)^2) + c \text{,}
\end{equation}
where $b$ originates in the imperfect interaction between data and read pulse and $c$ denotes the offset due to noise in the measurement.
Before using this model to fit the data presented in Fig.~4~(c) of the main text, one additional parameter is needed.
The energy of the control pulses is determined by measuring the average power on a powermeter and assuming all of this power is in the two control pulses.
In an experimental setup this can not be realized idealy, leading to a lowered power in the pulses, accompanied by a low noise floor in between pulses.
To account for this we add common fitting parameter $\eta$ and introduce this into the efficiency $\Theta = 1.92 \sqrt{\eta P_\mathrm{c}/\SI{}{\watt}}$.

The fit agrees well with the measured data (\cref{fig:MemoryFit}), yielding $\eta = \SI{0.39}{}$, $a = \SI{0.66}{}$, $b = \SI{0.51}{}$ and $c = \SI{0.05}{}$.
Using the duration of the two control pulses and their repetition rate of \SI{5}{\mega\hertz}, $\eta$ results in an extinction ratio of \SI{15.7}{\decibel} of the intensity modulator used to create the pulses.
In other words, during the measurement the noise floor is \SI{2.7}{\percent} of the peak power in the control pulses and contributes \SI{61}{\percent} to the average power measured on the powermeter.
Optimizing this to $\eta = \SI{1}{}$ would further lower the energy requirements of the system.

Note that increasing the control power further is not expected to increase the read-out efficiency significantly.
However optimizing the interaction, e.g. through pump pulse shortening \cite{zhu_stored_2007}, can increase the amplitudes $a$ and $b$ subsequently increasing the read-out efficiency.

\section{Optimization Potential of the Platform}
\begin{figure}
    \centering
    \includegraphics[width=0.8\textwidth]{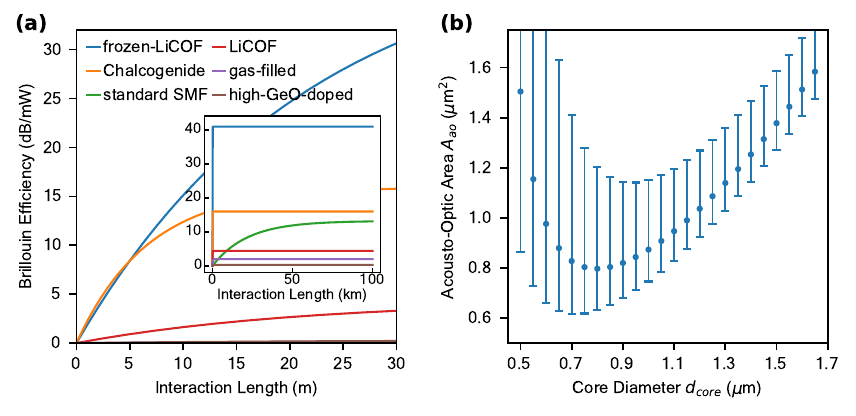}
    \caption{Future potential of the frozen LiCOF platform. \textbf{(a)} Theoretical efficiency (based on measured data in Fig.~2 of the main manuscript) for different fiber lengths limited by propagation losses. Inset shows km scale lengths. \textbf{(b)} $A_\mathrm{ao}$ of $HE_{1,1} \times L_{01}$ for measured refractive index with error bars based on the refractive index measurements errors.}
    \label{fig:Potential}
\end{figure}
While the measurements shown in the main text are done with a frozen LiCOF with $\SI{27.5}{\centi\meter}$ frozen length, freezing lengths on the order of meters is considered feasible.
For long lengths of fibers the total gain of the fiber is limited by the propagation losses of the material, typically accounted for by using the effective length \cite{kobyakov_stimulated_2010}
\begin{equation}
    L_\mathrm{eff} = \left( 1 - e^{-\alpha L} \right) / \alpha
\end{equation}
instead of $L$ when calculating the expected on-off gain.
For fibers with $L \ll 1/\alpha$ we obtain $L_\mathrm{eff} \rightarrow L$, while for longer fiber lengths we can not neglect the losses anymore and in the limit of $L \rightarrow \infty$ the effective length tends towards $1/\alpha$.
\Cref{fig:Potential}~(a) shows the evolution of the Brillouin efficiency, i.e. the total gain per milliwatt pump power, for increasing interaction lengths on the order of meters for different platforms.
The inset shows the same data but for kilometers of length, revealing how standard single-mode fibers outperform others in this regime due to their low losses.

A different approach of increasing the total gain of the platform is to increase $G_\mathrm{B}$ further.
As $g_\mathrm{B}$ is a material constant apart from $n_\mathrm{eff}$, this can mainly be done by decreasing $A_\mathrm{ao}$.
\Cref{fig:Potential}~(b) shows a simulation of $A_\mathrm{ao}$ of $HE_{1,1} \times L_{01}$ for a range of core diameters and the measured refractive index of solid \CS, showing that an improvement of $G_\mathrm{B}$ of roughly $\times \SI{1.5}{}$ is achievable by decreasing the core diameter to \SI{0.8}{\micro \meter}.

Finally, the coupling from SMF to LiCOF was not optimized for such small core diameters and can therefore be improved further by adapting the coupling fibers and splice procedure.
\end{document}